\documentclass{aa}  

\usepackage{graphicx}
\usepackage{txfonts}
\begin{document}

   \title{The Pristine Dwarf-Galaxy survey - VI. A VLT/FLAMES spectroscopic study of the dwarf galaxy Boötes~II}

   \author{Nicolas Longeard\inst{1,2}, Pascale Jablonka\inst{1,3}, Giuseppina Battaglia\inst{4,5}, Khyati Malhan\inst{6,7}, Nicolas F. Martin\inst{6,8}, Julio F. Navarro\inst{9}, Federico Sestito\inst{10}}
   \authorrunning{Nicolas Longeard\inst{1,2}}

   \institute{Laboratoire d'astrophysique, \'Ecole Polytechnique F\'ed\'erale de Lausanne (EPFL), Observatoire, 1290 Versoix, Switzerland   
    \and
    IFCA, Instituto de F\'isica de Cantabria (UC-CSIC), Av. de Los Castros s/n, 39005 Santander, Spain
    \and
    GEPI, Observatoire de Paris, Universit\'e PSL, CNRS, Place Jules Janssen, F-92195 Meudon, France
    \and
    Instituto de Astrofísica de Canarias, Calle Vía Láctea s/n, 38206 La Laguna, Santa Cruz de Tenerife, Spain
    \and
    Universidad de La Laguna, Avda. Astrofísico Francisco Sánchez, 38205 La Laguna, Santa Cruz de Tenerife, Spain
    \and
    Max-Planck-Institut f¨ur Astronomie, K¨onigstuhl 17, D-69117, Heidelberg, Germany
    \and
    Myrspoven AB, V¨astg¨otagatan 1, 11827 Stockholm, Sweden
    \and
    Université de Strasbourg, CNRS, Observatoire astronomique de Strasbourg, UMR 7550, F-67000 Strasbourg, France
    \and
    Dept. of Physics and Astronomy, University of Victoria, P.O. Box 3055, STN CSC, Victoria BC V8W 3P6, Canada
    \and
    Centre for Astrophysics Research, Department of Physics, Astronomy and Mathematics, University of Hertfordshire, Hatfield, AL10 9AB, UK
   }


 
  \abstract
   {The Milky Way has a large population of dwarf galaxy satellites. Their properties are sensitive to both cosmology and the physical processes underlying galaxy formation, but these properties are still not properly characterised for the entire satellite population.}
   {We aim to provide the most accurate systemic dynamical and metallicity properties of the dwarf galaxy Boötes~II (Boo~II).}
   {We use a new spectroscopic sample of 39 stars in the field of Boo~II (heliocentric distance of $\sim 66$ kpc) with data from the Fiber Large Array Multi Element Spectrograph (FLAMES) mounted on the Very Large Telescope (VLT). The target selection is based on a combination of broadband photometry, proper motions from Gaia, and the metallicity-sensitive narrow-band photometry from the Pristine survey that is ideal for removing obvious Milky Way contaminants.}
   {We found 9 new members, including 5 also found by the recent work of \citet{Bruce23}, and the farthest member to date (5.7 half-light radii from Boo~II centroid), extending the spectroscopic spatial coverage of this system. Our metallicity measurements based on the Calcium triplet lines leads to the detection of the two first 2 Extremely Metal-poor stars ([Fe/H] $< -3.0$) in Boo~II. Combining this new dataset with literature data refines Boo~II's velocity dispersion ($5.6^{+1.8}_{-1.1}$ km s$^{-1}$), systemic velocity ($-126.8^{+2.0}_{-1.5}$ km s$^{-1}$) and shows that it does not show any sign of a significant velocity gradient (d$\langle v \rangle$/d$\chi = 0.6 ^{+0.6}_{-0.4}$  km s$^{-1}$ arcmin$^{-1}$, or $-0.5/1.9$ km s$^{-1}$ arcmin$^{-1}$ as $3\sigma$ upper limits). We are thus able to confirm the kinematic and metallicity properties of the satellite as well as identify new members for future high-resolution analyses.}


   \maketitle

\section{Introduction}

The successive large coverage photometric surveys over the last two decades, from the Sloan Digital Sky Survey \citep[SDSS]{york2000}, the Panoramic Survey Telescope And Rapid Response System \citep[PS1]{chambers16}, or the Dark Energy Survey \citep[DES]{des05}, to the upcoming Legacy Survey of Space and Time \citep[LSST]{ivezic19}, have led to the discovery of dozens of faint galaxy companions orbiting the Milky Way (MW). The faintest ones are usually referred to as Ultra-Faint dwarf galaxies (UFD).

The current UFD population of the MW has been extensively studied. Their morphological properties (morphology, luminosity, mass) as well as their spectroscopic properties (metallicity, chemical enrichment, orbit) were an immediate focus point of the community. Most UFD studies have a singular goal: to compare their observed morphology with the ones predicted by the various numerical simulations made throughout the years (e.g. \citealt{sawala16} or \citealt{read19} for the mass, \citealt{revaz23} for the morphology, \citealt{sanati23} for the stellar abundances, \citealt{sanati24} for the mass and stellar population). In doing so, we are able to improve our understanding not only of the cosmological significance of UFDs (e.g. \citealt{springel08}, \citealt{read19}, \citealt{revaz23}) but also of the physical processes linked to the formation and evolution of galaxies, such as stellar feedback (e.g. \citealt{agertz2020}, \citealt{sanati23}).

Therefore, a large number of spectroscopic studies of UFDs have been carried out over the years. Until recently, the vast majority of these studies from different teams with different observational strategies using different observing facilities have had but one main goal: derive the intrinsic properties of the satellite (morphology, dynamical mass, stellar abundances). To do so, it is imperative to derive the membership of as many UFD stars as possible. A task made challenging by the foreground contamination of the MW stars. 

UFDs are usually approximated as  "simple" systems characterized by the mean and dispersion of their velocity and metallicity distributions, which are usually considered enough to offer a comprehensive view of their properties (\citealt{simon19} and references therein). This simplistic view of the UFDs' stellar population is now more and more disputed with the rise of observing strategies specifically designed to more efficiently weed out the MW stellar contamination. This also allows to find more members with larger spatial coverage than before. These strategies essentially rest on the use of specific photometric bands that are able to trace the metal content of stars and to discard the obvious metal-rich population that should not inhabit UFDs (e.g. \citealt{longeard21}, \citealt{longeard22} for the Pristine survey, \citealt{chiti21} for SkyMapper, or \citealt{pan24} with DECam photometry). These studies indeed point towards a more complex picture for both their dynamical (i.e. the mass of all components of the galaxy, see \citealt{wolf10} and \citealt{errani18}) and metallicity properties. In that sense, such spectroscopic studies are needed to paint the most accurate picture of the overall UFD population which is critical to properly constrain our cosmological and galaxy models by comparing observations with simulations.

The following work is directly related to this effort: we spectroscopically observed the UFD Boo~II to enlarge the catalog of member stars and to spatially extend this catalog.
Boo~II has already been the subject of two past spectroscopic studies, \citet[K09]{koch09} and \citet[B23]{Bruce23}, with the latter published during the completion of this manuscript, after our observations had already been carried out and analysed. K09 used Gemini/GMOS multi-object spectroscopy \citep{hook04} to find the first 5 members in the system and derived a systemic velocity of $117 \pm 5.2$ km s$^{-1}$ and a large but highly uncertain velocity dispersion of $10.5 \pm 7.4$ km s$^{-1}$. Their metallicity was found to be of $-1.79 \pm 0.05$. Recently, B23 found 9 new member stars and confirmed the membership of 3 stars of K09 using Magellan/IMACS spectroscopy \citep{dressler11}. In particular, B23 showed that the mean quantitative results of K09 were heavily biased with an updated velocity of $-130.4 ^{+1.4}_{-1.1}$ km s$^{-1}$ and a metallicity of $-2.71 ^{+0.11}_{-0.10}$, most likely due to the observational setup used by K09 that is not ideal to derive accurate velocities although it is enough for membership inference. B23 also tightly constrained the velocity dispersion ($2.9^{+1.6}_{-1.2}$ km s$^{-1}$) and therefore the mass-to-light ratio of Boo~II ($460^{+1000}_{-440}$ M$_\odot$ L$_\odot^{-1}$). Their main results are summarized in Table 1.

The following introduces our brand-new Boo~II observations and subsequent dynamical and metallicity analyses. Section 2 details the data selection, observations and reduction procedures. Section 3 presents our membership, dynamical and metallicity results. Finally, section 4 summarizes our findings and discusses our current knowledge of Boo~II in the light of all the studies existing on this UFD.

\begin{table}
\centering
\begin{tabular}{|llll|} 

 \hline
Property & Lit. value & Reference & This work  \\ 
  \hline
$d_\mathrm{GC}$ (kpc) & $42 \pm 1 $ & (1) & $-$ \\ 
$r_h$ (') & $2.60 \pm 0.8$ & (1), (2) & $-$  \\
$r_h$ (pc) & $39 \pm 5$ & (1), (2) & $-$  \\
$\langle v \rangle$ (km s$^{-1}$) & $-130.4^{+1.4}_{-1.1}$ & (3), (4) & $-126^{+2.0}_{-1.5}$ \\
$\sigma_\mathrm{v} $ (km s$^{-1}$) & $2.9^{+1.6}_{-1.2}$ & (3), (4) & $5.6^{+1.8}_{-1.1}$ \\
$\mathrm{[Fe/H]}$ & $-2.71 ^{+0.06}_{-0.10}$ &  (2), (3), (4), (5) & $-2.3 \pm 0.2$  \\

 \hline
\end{tabular}
\caption{Summary of Boo~II' prior literature (K09 and B23) properties. The reference numbers correspond to the following list: (1) \citet{munoz18}, (2) \citet{walsh08}, (3) \citet{koch09}, (4) \citet{Bruce23}, (5) \citet{ji16}}
\label{table:1}
\end{table}

\section{Spectroscopic observations}

\begin{figure*}
\begin{center}
\centerline{\includegraphics[width=\hsize]{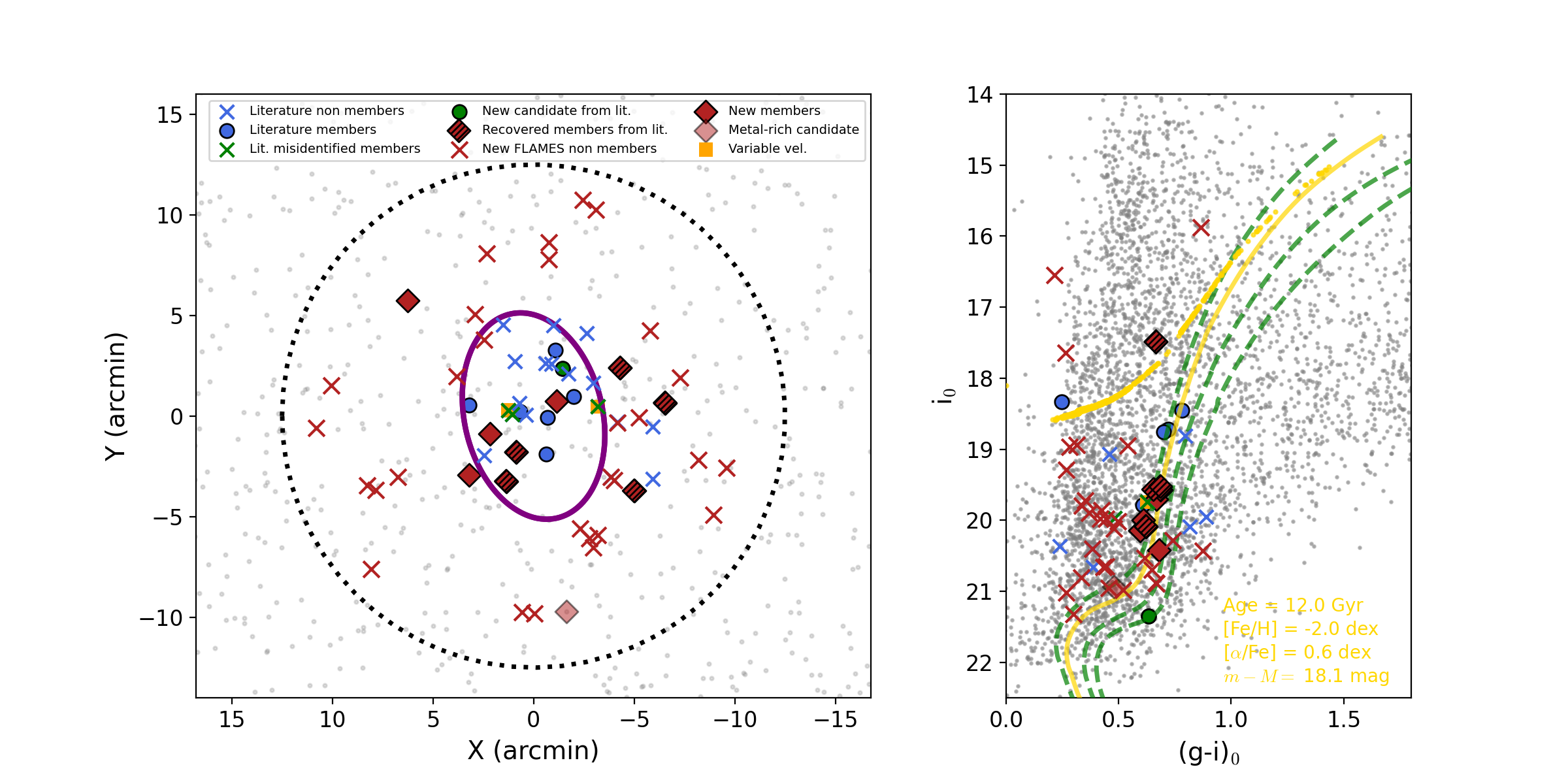}}
\caption{\textit{Left panel:} Spatial distribution of the AAT spectroscopic sample. Newly discovered members are shown as red diamonds. Non-members from the AAT sample are shown as red crosses. Previously known members from the literature (K09 + B23) are represented as smaller blue circles. Misidentified literature members are shown as green crosses. The two half-light radii of Boo~II as inferred by \citet[M18]{munoz18} are shown as a black ellipse. The small grey dots show all stars in the field with good quality SDSS photometry (i.e. broadband photometry uncertainties below $0.2$ mag). \textit{Right panel:} CMD of our spectroscopic sample superimposed with the favoured Boo~II Dartmouth isochrone in yellow, chosen to match the system's spectroscopically identified RGB stars. Three other isochrones of varying metallicity (one more metal-poor at [Fe/H] $\sim -2.5$, two more metal-rich at $\sim -1.5$ and $\sim -1.2$) are also overplotted as dashed green lines. The g and i magnitudes are from SDSS.}
\label{field} 
\end{center}
\end{figure*}

This section provides details on the target selection, observations and data reduction first, then briefly explains our velocity and equivalent widths (EWs) derivation method.

\subsection{Data selection}

All targets were selected based on the colour-magnitude diagram (CMD), narrow-band photometry, and with Gaia astrometry, as detailed in the following subsections. 

\subsubsection{The Pristine survey selection}
Pristine is a photometric survey (\citealt{starkenburg17}, \citealt{martin23}) relying on a narrow-band, metallicity-sensitive photometry centered on the Calcium H\&K doublet lines from the Canadian France Hawaii Telescope \citep[CFHT]{boulade03}. It is successful at finding metal-poor stars against the more metal-rich MW contamination (\citealt{youakim17}, \citealt{aguado19}, \citealt{arentsen20}) and is therefore particularly suited for the UFDs metal-poor population (\citealt{longeard20}, \citealt{longeard21}, \citealt{longeard22}). To derive the Pristine photometric metallicities, the procedure of \citet{starkenburg17} is followed and we refer the reader to their section 3 for details. The main difference with \citet{starkenburg17} is that the Pristine survey footprint has since been greatly expanded. Therefore, our training sample to create the $CaHK$ to [Fe/H] model is larger than theirs. 

To summarize the process, Pristine observes stars with a narrow-band filter centered on the $CaHK$ doublet lines. These stars also have a counterpart in SDSS to obtain classical, broadband photometry. To build the $CaHK$ to [Fe/H] model, we restrict ourselves to stars observed with Pristine that also have an SDSS spectroscopic metallicity measurement. This training sample is cleaned from potential white dwarfs, mediocre $CaHK$ and broadband photometry following the criteria of \citet{starkenburg17}. The remaining stars are then placed in a $g-i$, $CaHK - g - 1.5\times(g-i)$ colour-colour diagram. In this a colour-colour space, they naturally separate as a function of their metallicity (Figure 11 of \citet{starkenburg17}). By binning this colour-colour space and using the stars' SDSS metallicities as a reference, we are able to calibrate this space and derive the photometric metallicity of any star with SDSS broadband and Pristine photometries. 

\subsubsection{Selection criteria}

Three main criteria were applied to select potential members for our spectroscopic observations:

\begin{itemize}
\item Stars located further than 0.2 mag from the best-matching Boo~II isochrone (Age = 13 Gyr, [Fe/H] = -2.4, [$\alpha$/Fe] = 0.0, $m - M$ = 18.10) from the Darmouth library \citep{dotter08} were discarded. The distance modulus was taken from \citet{munoz18}.
\item Their location on the Pristine colour-colour diagram must be located above the region occupied by the population of UFDs \citep{longeard23}, indicating a metallicity lower than that of the vast majority of MW stars. This roughly corresponds to metallicities lower than $-1.0$ dex.
\item The proper motion membership probability of all targets must be at least $1$\%, based on the Gaia Data Release 3 \citep{gaia_dr3}. These membership probabilities are computed assuming two multivariate gaussian populations in proper motion space, for Boo~II and the MW, respectively, based on the systemic proper motion of \citet{battaglia22} and \citet{mcconnachie_venn20}.
\end{itemize} 

\subsection{Data acquisition}

The spectroscopic sample was obtained using the Very Large Telescope \citep[VLT]{pasquini02} and its FLAMES multi-object spectrograph. The HR21 grating was used to encompass the calcium triplet (CaT) lines, with a spectral resolution R of $\sim 18000$. The observations were carried out throughout 2 semesters, from April 2022 to February 2023, observing 9 sub-exposures of 2775 seconds each. Only the ninth sub-exposure were observed in February 2023, the rest being observed in April/May 2022. Unfortunately, this prevents us from from performing a robust binary test between the sub-exposures carried out in 2022 and the one in 2023, since it requires to analyze the ninth exposure alone, which considerably reduces the SNR of each spectrum. However, this was performed when possible and the velocities between the 2022 and 2023 observations were consistent within their respective uncertainties, indicating that no obvious binaries might affect the results.

All sub-exposures are indicated as good-quality observations, i.e. graded A (8) or B (1) by the ESO observing team. To summarize, when requesting an observation at ESO, observational constraints are also provided. A grade A corresponds to the situation where all constraints are met, while B is for observations for which they were not all met, but were within 10\% of the requested values. The field is centered on the system based on the coordinates of \citet{munoz18}. This field is shown in Figure \ref{field} and extends as far as $\sim 8$ half-light radii (r$_h$) of Boötes~II (Boo~II). 

The few unassigned fibers left were filled with even lower priority stars and interesting, potentially extremely metal-poor (EMP, [Fe/H] $< -3.0$) MW halo stars in the field according to Pristine.

\subsection{Data reduction}

\begin{figure}
\begin{center}
\centerline{\includegraphics[width=1.1\hsize]{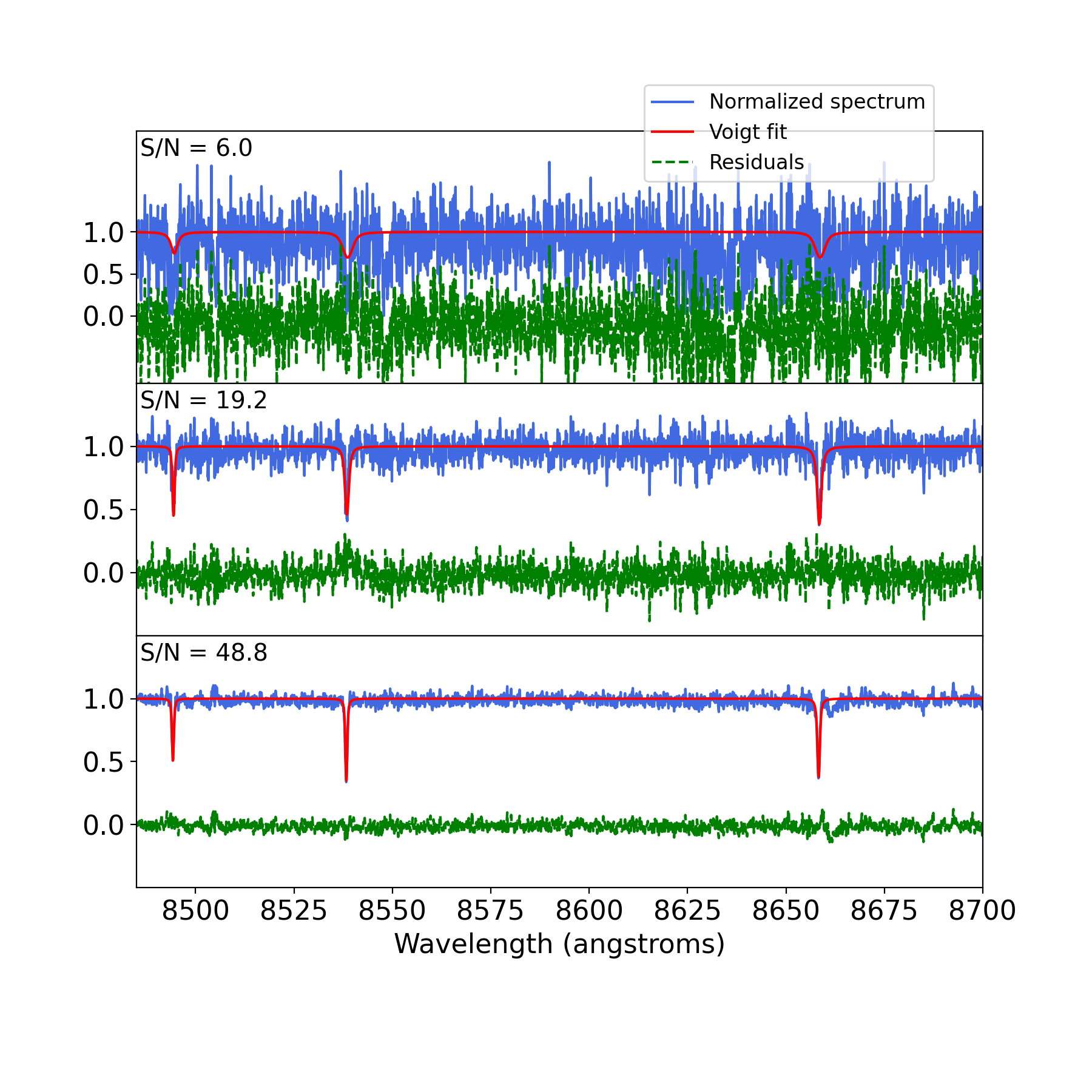}}
\caption{Example spectra of three candidate member stars in our FLAMES dataset centered on the CaT lines. The low, mid and high S/N regimes are represented here. The normalised spectra are shown with solid blue lines while the fits derived from our pipeline for Voigt profiles are shown with solid red lines. Residuals in the Voigt cases are shown for each case below the spectra as green dashed lines. These stars have a heliocentric velocity of $ -133.2 \pm 1.5$ (the "metal-rich" candidate in Figure 2), $-111.3 \pm 0.6$ and $-133.3 \pm 0.2$ km s$^{-1}$ from top to bottom.}
\label{spectra} 
\end{center}
\end{figure}

\begin{figure}
\begin{center}
\centerline{\includegraphics[width=1.1\hsize]{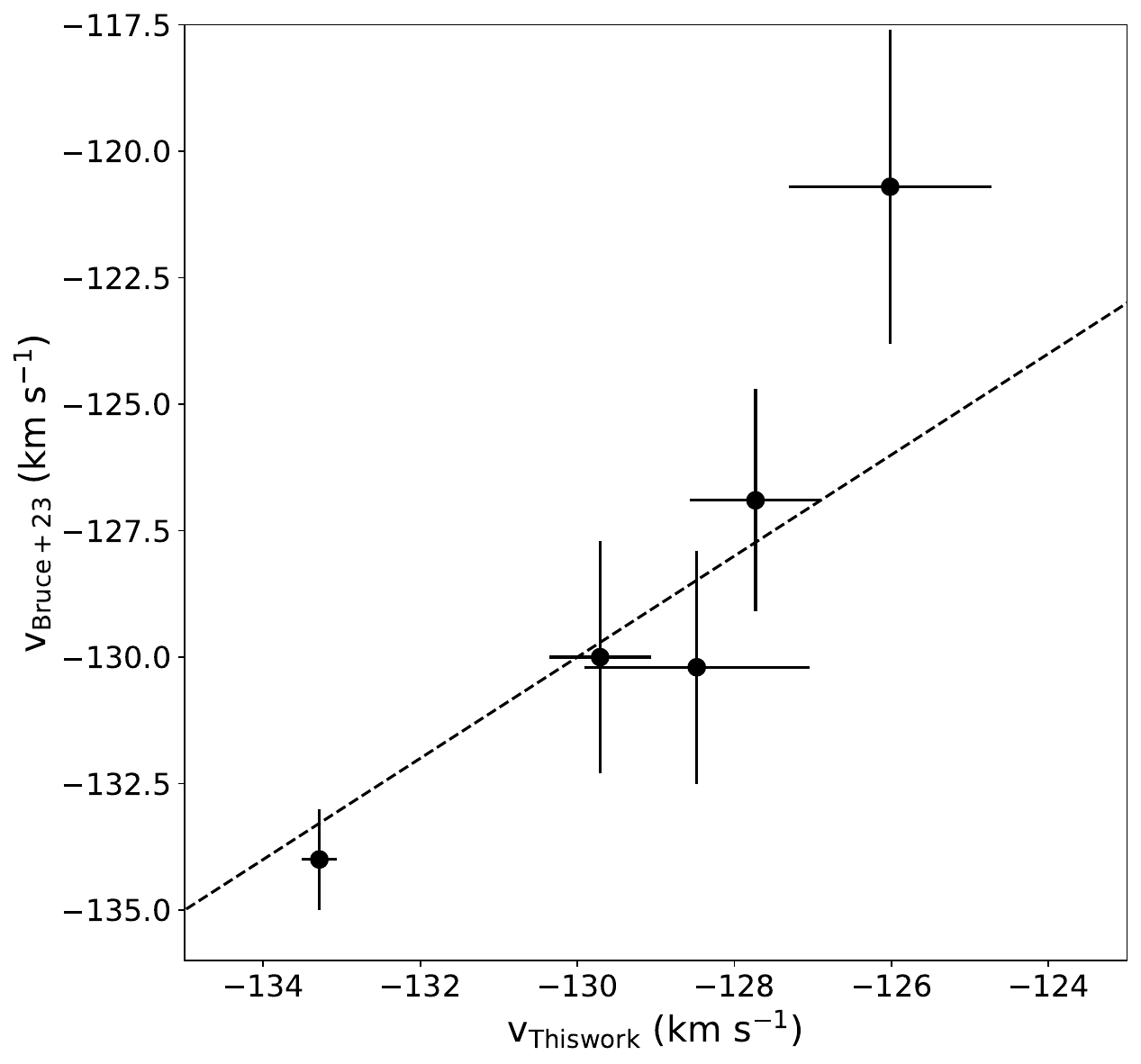}}
\caption{Comparison between the heliocentric velocity of B23 and the ones found in this work for the members stars in common.}.
\label{comparison} 
\end{center}
\end{figure}

The standard ESO package to reduce GIRAFFE data \citep{melo09} was used to reduce the data, without any modification to the pipeline. 

Three examples of spectra for low (6.0), mid (15.0) and high (164) signal-to-noise (S/N) ratios per pixel are shown in Figure \ref{spectra}. Each spectrum was carefully visually inspected and discarded if its quality was too poor to obtain a proper fit of any of the three CaT lines. This mosty involves low S/N spectra bellow $3$, but also mid S/N ones for which the CaT lines are too contaminated (from e.g. sky substraction) to be properly fitted. This step led to the rejection of 5 spectra.

We normalize the spectra following the method of \citet{battaglia08}, i.e. through an iterative k-sigma clipping non-linear filter. The heliocentric velocities and equivalent widths (EWs) of each spectrum are then obtained using our in-house pipeline described in detail in \citet{longeard22}, which has already been extensively tested against known metallicities and velocities and recently used by \citet{longeard23}. Each CaT line is modeled with a Gaussian and Voigt profiles and their position is found by minimizing the squared difference between a synthetic spectrum composed of three Gaussian/Voigt profiles and the observed spectrum. The EWs are calculated by integrating the best fit around each line in a 15\AA $\;$ window. This is performed with a Monte Carlo Markov Chain \citep{hastings70} algorithm with a million iterations per spectrum. The MCMC produces posterior probability functions that allow us to derive the uncertainty on each parameter.

The median of the velocity uncertainty is $1.2$ km s$^{-1}$ for the new FLAMES sample, $1.7$ km s$^{-1}$ for entire sample (literature + FLAMES, details in section 3.2) and $1.3$ km s$^{-1}$ when the sample is restricted to Boo~II literature member stars only. Note that for the rest of this work, the "literature" refers to the two previous spectroscopic analysis of Boo~II, i.e. B23 and K09. Figure \ref{comparison} shows that our velocities agree very well with those of B23 for five member stars in common, and that FLAMES data reduce the velocity uncertainties by a factor of $\sim 2$ on average.

The UFDs Leo~IV and Leo~V were also observed during this program and subjected to the same data treatment. However, the lack of enough new members in our data explains that we do not analyse those data in more details in the following work.

\section{Results}

We present in this section the results of our analysis. The metallicity results will be presented first since they are needed to derive BooII's kinematic properties.

\subsection{Metallicity properties}

\begin{figure}
\begin{center}
\centerline{\includegraphics[width=1.1\hsize]{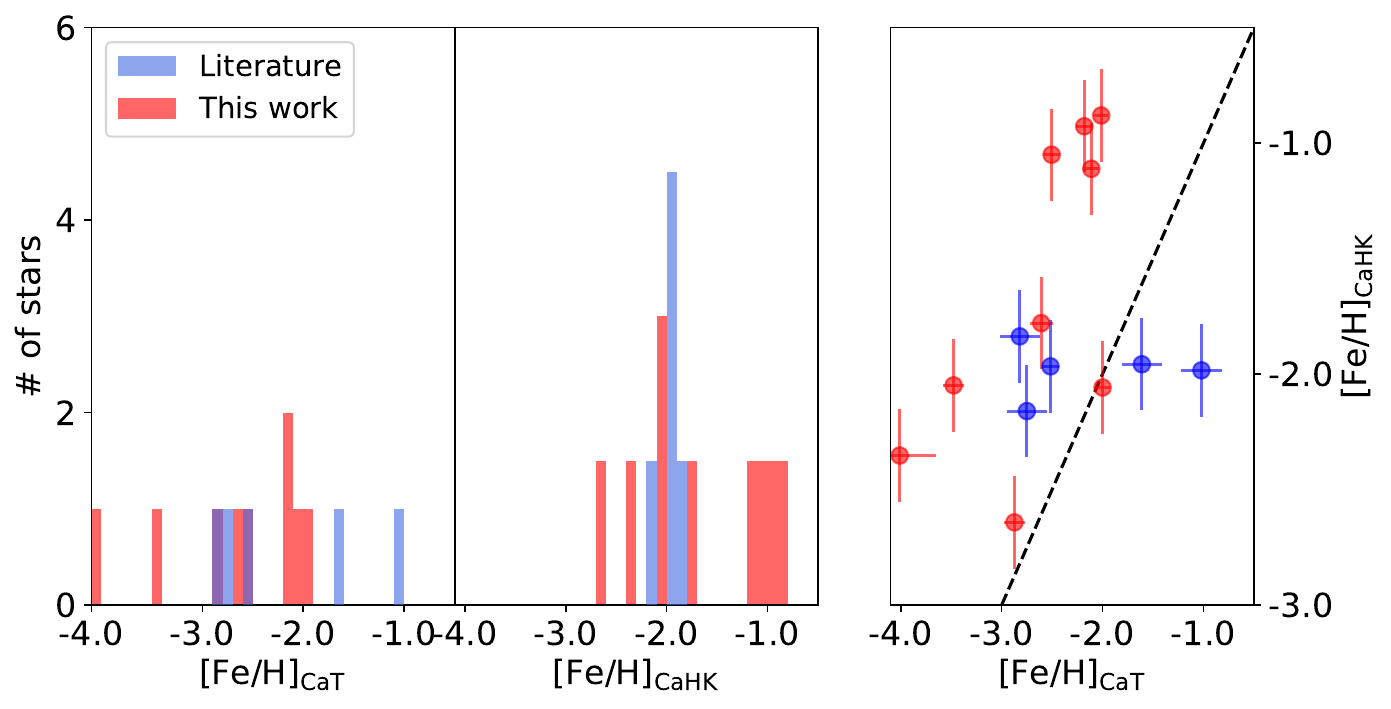}}
\caption{Metallicity distribution functions (MDFs) for the CaT (left) and $CaHK$ (center) cases. The right panel shows the comparison between the two metallicities, with the 1:1 line as the black dashed line.}
\label{MDFs} 
\end{center}
\end{figure}

\begin{figure}
\begin{center}
\centerline{\includegraphics[width=1.1\hsize]{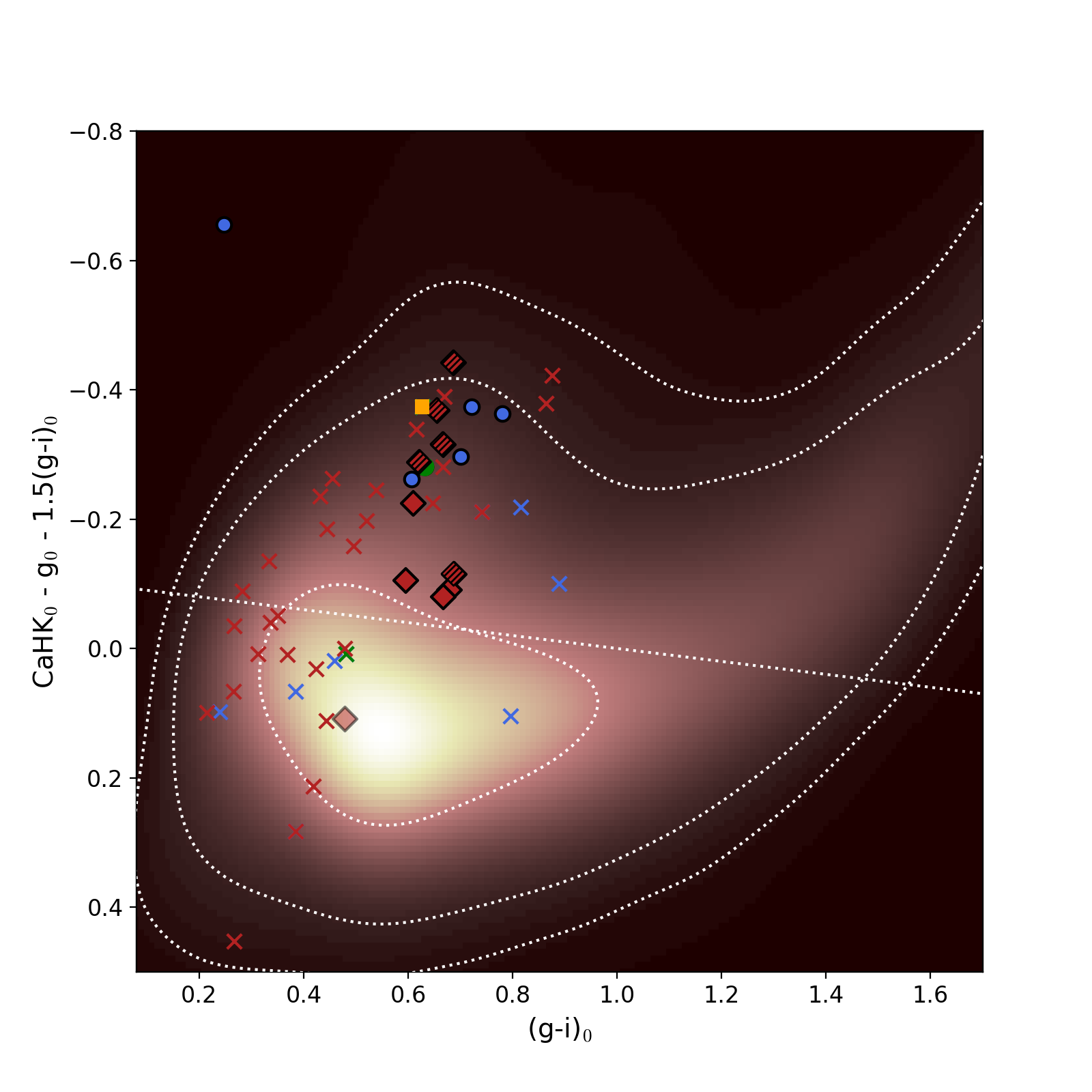}}
\caption{$CaHK$ colour-colour diagram, with the temperature proxy ($g-i$)$_{0}$ on the x-axis and the Pristine colour containing the metallicity-sensitive information. The density map in the background of the plot is produced using MW stars in the field of Boo~II, with 1,2 and 3$\sigma$ contours shown as dotted white lines. In this diagram, the metallicity goes down as the y-axis colour decreases. The dashed straight line shows the cut applied to the data to distinguish likely the metal-poor from the likely metal-rich stars. The colours and markers used are the same as those of Fig. \ref{field}.}
\label{cahk} 
\end{center}
\end{figure}

The metallicities of the stars in our sample are derived in two ways that will be presented below: 1) using spectra when the S/N is greater than 10 for the FLAMES and literature members, and 2) using the Pristine narrow-band, metallicity-sensitive data. For the latter, a calibration of the $CaHK$ space for giant stars is used.

To derive the spectroscopic metallicities, we use the same process as in \citet{longeard22} and \citet{longeard23}, i.e. derive the EWs of the three CaT lines, then use the calibration of \citet{carrera13} that translates them into a metallicity value with an associated uncertainty that takes into account the following sources: uncertainty on the EWs, the distance modulus, the photometry and the calibration's coefficients. The results are reported in Table 2 and shown in Figure \ref{MDFs}. The left panel shows that we identify two members as EMPs, including one at the edge of the calibration range of \citet{carrera13} at $\sim -4.0$. The right panel shows that a bias exists between $CaHK$ and spectroscopic metallicities of the order of $\sim 0.5$ dex. We do not understand the source of this discrepancy, though it most likely stems from different photometric zero-points between all CCDs in the field. However, this does not impact our results since our Pristine-based selection does not rely on the metallicity value itself but on their location of the Pristine colour-colour diagram shown in Figure \ref{cahk}, and that our selection is quite generous on that colour space.

In the rest of this study, and coming from experience, stars with a $CaHK$ uncertainty above 0.2 mag often yields mediocre/very uncertain photometric metallicities and their metallicities are not computed. The results are shown in the right panel of Figure \ref{MDFs}. The right panel clearly shows a metallicity peak at around $-2.0$ that corresponds to Boo~II's population.

\subsection{Analysis}

\begin{figure}
\begin{center}
\centerline{\includegraphics[width=1.1\hsize]{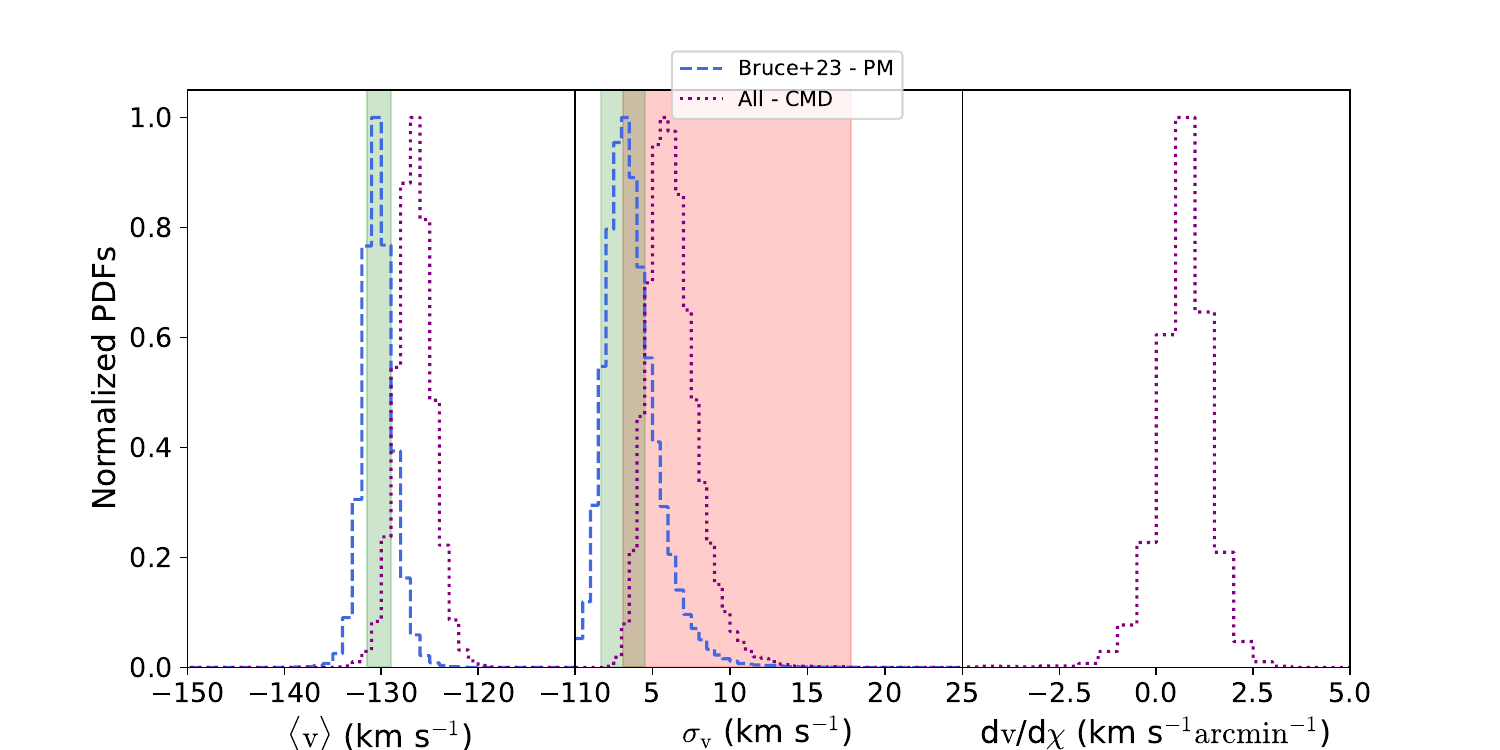}}
\caption{Posterior PDFs of the main dynamical properties of Boo~II, i.e. the systemic velocity (\textit{left panel}), the intrinsic velocity dispersion (\textit{middle panel}) and velocity gradient (\textit{right panel}). The blue dashed PDFs show the results of our analysis using only the B23 sample. The purple PDF shows the result of this work using the entire sample (i.e. K09 + B23 + FLAMES). The shaded area indicates the 1$\sigma$ interval inference of K09 (green) and B23 (red). This K09 interval is not shown for the systemic velocity as their velocity was strongly biased ($\sim -110$ km s$^{-1}$, and therefore outside the plot.)}
\label{pdfs_dynamics} 
\end{center}
\end{figure}

\begin{figure}
\begin{center}
\centerline{\includegraphics[width=1.1\hsize]{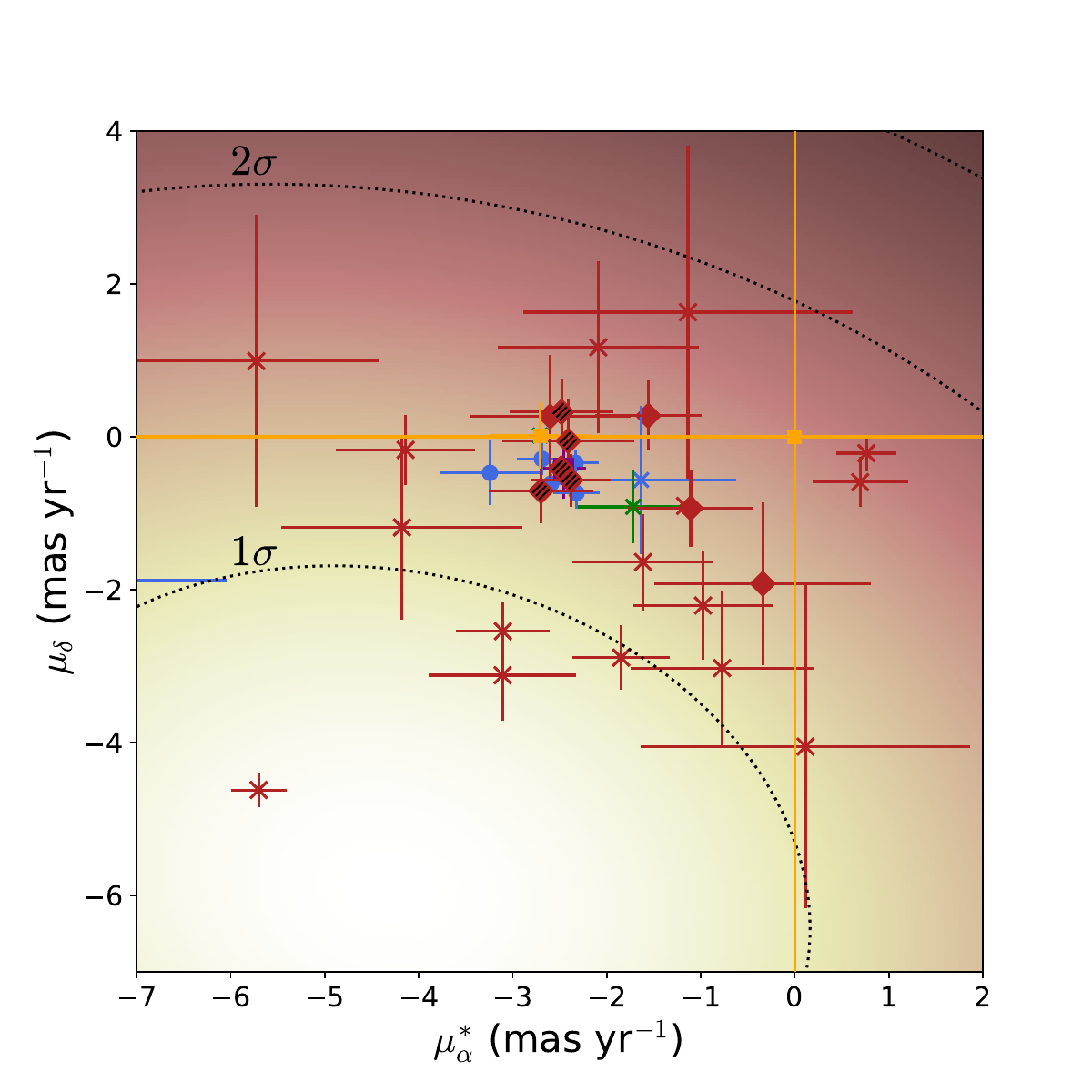}}
\caption{Proper motions of the full spectroscopic sample. The plot is restricted to an area around the proper motion of Boo~II for the sake of visibility, but other stars (clearly non-members) are located outside this region. The density background shows the density of MW star in the field of Boo~II, with $1$ and $2\sigma$ contours shown as black dashed lines. The colour and marker schemes are the same as in previous plots.}.
\label{pm} 
\end{center}
\end{figure}

\begin{figure*}
\begin{center}
\centerline{\includegraphics[width=1.1\hsize]{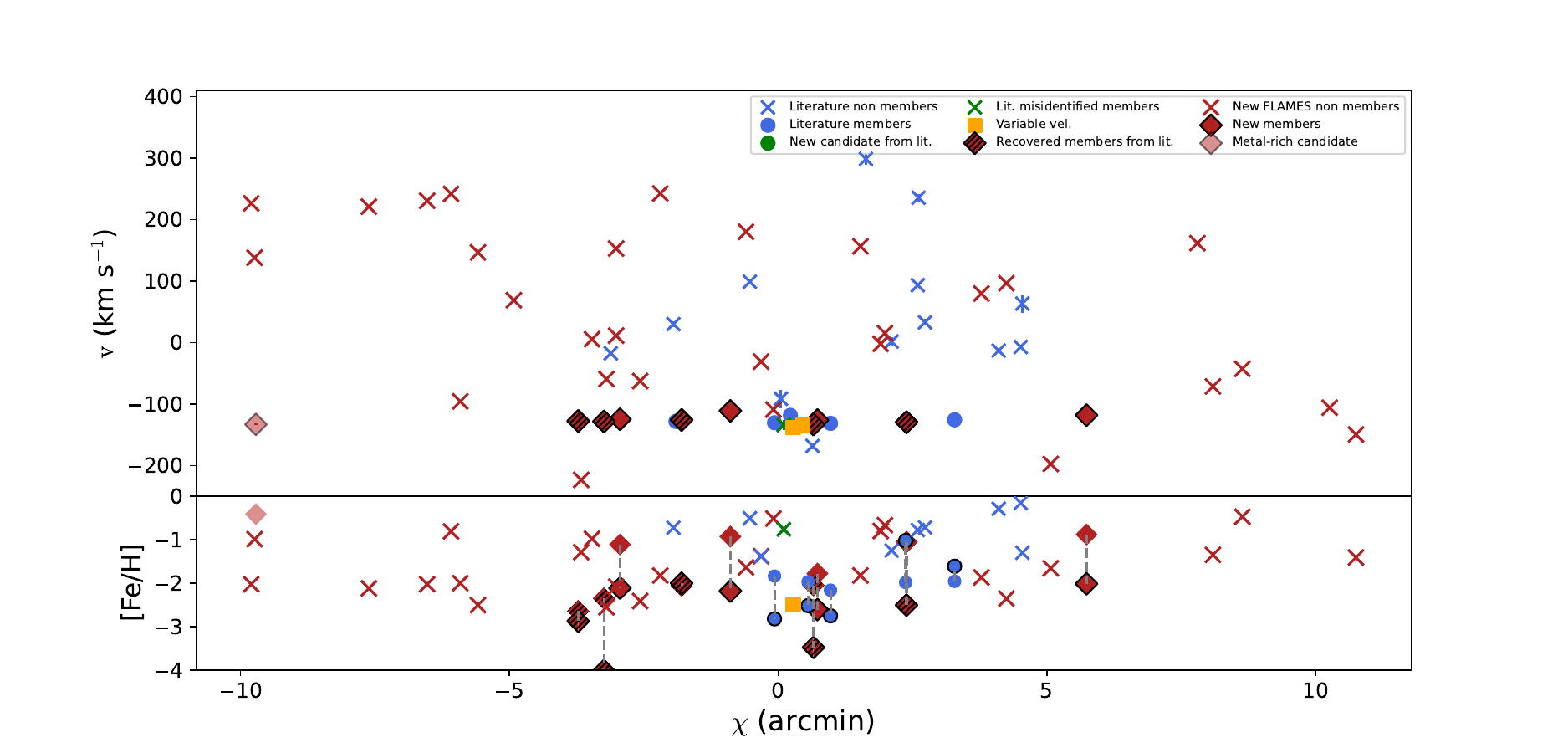}}
\caption{Summary of the velocity and metallicity measurements of the spectroscopic sample to investigate potential velocity/metallicity gradients. The velocity uncertainties are reported in the plot but are so small compared to the scale of the y-axis that they are overall not visible, except for a few cases with extremely large uncertainties. \textit{Top panel:} Velocity vs. position along Boo~II's major axis. The colour and marker schemes are similar to what was shown in the previous plots. The literature (i.e. K09 + B23) and new members do not display any sign of a velocity gradient, even by including the potential more metal-rich member candidate at $\chi \sim -10$ arcmin. \textit{Lower panel:} CaT (with black contours) and CaHK (without contours) metallicities. In the case where both are available for the same star, the two values are linked by a grey-dashed line.}
\label{v_vs_r} 
\end{center}
\end{figure*}

To derive the dynamical properties of the system, we add more criteria to clean the sample besides the ones detailed in section 2.1.2. These criteria are:

\begin{itemize}
    \item Their ratio of parallax over parallax uncertainty taken from Gaia must be lower than 2.0.
    \item They must not have been identified as potentially variable-velocity stars in the literature.
\end{itemize}

Once all these criteria are applied, the final sample consists of 39 stars. Given the number of fibers that were still available after the cleaning of section 2.1.2, we decided to observe probable MW halo stars flagged as metal-poor by the $CaHK$ photometry as target of opportunities, even though they are not linked to the analysis and therefore not included in the 39 stars in the final sample.

The dynamical properties of Boo~II are derived following the formalism of \citet{martin_jin10} combined with the likelihoods described in their Equations (2) and (3):

\begin{equation}
\begin{aligned}
\mathcal{L}(\mathrm{v}_{\mathrm{r},k}, \delta_{\mathrm{v},k} | \langle \mathrm{v}_{\mathrm{Boo}} \rangle, \langle \mathrm{v}_{\mathrm{MW}} \rangle, \sigma^\mathrm{Boo}_{\mathrm{v}}, \sigma^\mathrm{MW}_{\mathrm{v}}, d\mathrm{v}/d\chi, \theta, \eta_\mathrm{Boo}) = \\
\prod_k \;  \bigg[ \eta_\mathrm{Boo} (\frac{1}{\sqrt{2 \pi \sigma}}) \times \mathrm{exp}(\frac{1}{2}\Delta_\mathrm{v}/\sigma^{2}) \mathcal{L}^{\mathrm{Boo}}_{\mathrm{PM}}+ \\
(1 - \eta_\mathrm{Boo}) \mathcal{G}(\mathrm{v}_{\mathrm{r},k}, \delta_{\mathrm{v},k},\langle \mathrm{v}_{\mathrm{MW}} \rangle, \sigma^\mathrm{MW}_{\mathrm{v}}) \mathcal{L}^{\mathrm{MW}}_{\mathrm{PM}} \bigg],
\end{aligned}
\end{equation}

\noindent We define $\Delta_\mathrm{v}$ such that $\Delta_\mathrm{v} = \mathrm{v}_{\mathrm{r},k} - y \times d\mathrm{v}/d\chi + \langle \mathrm{v}_{\mathrm{BooII}} \rangle$ with d$\mathrm{v}$/d$\chi$ the systemic heliocentric velocity gradient, and $\chi$ the galacto-centric distance along the position angle $\theta$. $y$ is the angular distance computed such that $y_k = X_k\sin{\theta} + Y_k\cos{\theta}$ and $\theta$ the direction of the velocity gradient. We also define $\sigma = \sqrt{(\sigma^\mathrm{BooII}_\mathrm{v})^{2} + (\delta^k_\mathrm{v})^{2}}$, with $\sigma^\mathrm{BooII}_\mathrm{v}$ the intrinsic Boo~II velocity dispersion and $\delta^k_\mathrm{v}$ the individual velocity uncertainty of each star. Finally, $\eta_\mathrm{BooII}$ is the Boo~II member fraction of the spectroscopic sample. \\

The final posterior probability distribution functions (PDFs) of Boo~II's dynamical properties are shown in Figure \ref{pdfs_dynamics}. This figure illustrates that our results are compatible with the ones of B23. To check the validity of our code, we first derived the systemic velocity and velocity dispersion of Boo~II only using the dataset of B23 and found the blue dashed PDF. This PDF falls perfectly onto the green shaded area corresponding to the results reported by B23. This plot shows that Boo~II does not show clear evidence of a velocity gradient.

From this analysis, we can derive the membership probability of all stars in the sample. To classify a star as a member, we add a 10\% dynamical membership probability, computed from Equation 1, to the criteria detailed in section 2.1.2 and the beginning of section 3.2.

Nine new members are found in the new FLAMES dataset, including 6 also identified by B23. Furthermore, the membership status of one other star is debatable as it is located right in the MW region of the Pristine colour-colour diagram, with no proper motion information to help in the decision-making process (Figure \ref{pm}) Its CMD location, bluer than the mean Boo~II population in Figure \ref{field}, suggests a lower metallicity than that of the UFD. Furthermore, one star classified as a non-member in the literature is found to be a tentative member in this study. It is shown as a green circle in Figures \ref{field}, \ref{cahk} and \ref{pm}. Its velocity and photometric metallicity are perfectly compatible with Boo~II. However, it does not have a PM measurement, and its CMD location is $\sim 0.15$ mag redder than that of the mean Boo~II population, which does not allow us to be certain of its membership. Three likely misidentified members from the literature, shown as green crosses, are also identified. Their properties are shown in Table 3. Finally, the different membership probability components are detailed in Table 4.

The entire dataset's velocity, metallicity and position along Boo~II's major-axis are summarized in Figure \ref{v_vs_r}. This plot is also useful to investigate any significant velocity/metallicity gradient along Boo~II major axis. The visual impression confirms our quantitative analysis below in that there does not seem to be any radial trend on any of the two properties. The lower panel of this plot also illustrates once more that the Pristine metallicities are overestimated when compared to their spectroscopic counterparts, when available.

The recent work of \citet{pan24} derives photometric metallicities and therefore promising candidates for Boo~II from their DECam photometry. Their table of candidates contains two (RA = 209.49804167$\deg$, DEC = 12.86472222$\deg$ --- RA = 209.5725$\deg$, DEC = 12.8035$\deg$) of our new members (i.e. not the ones in common with B23). Both are classified as "candidate with low purity due to their faintness" by \citet{pan24}. \\

Finally, we find a systemic velocity ($-126.8^{+2.0}_{-1.5}$ km s$^{-1}$) and velocity dispersion $5.6^{+1.8}_{-1.1}$ km s$^{-1}$) slightly larger, but statistically compatible with B23 at the $1\sigma$ level. The existence of a velocity gradient is investigated for the first time, and this investigation shows that Boo~II shows no convincing sign of a significant velocity gradient: it is consistent with the null hypothesis at the $1.5\sigma$ level ($0.6 ^{+0.6}_{-0.4}$ km s$^{-1}$). This implies that the Boo~II mass is not biased by tidal interactions or any unexpected internal velocity distribution.

\section{Summary and conclusion}

We present new spectroscopic observations  with the FLAMES/VLT spectrograph, of  candidate member stars in and around the faint UFD Boo~II.  Our initial selection was based on the combination of the Pristine 
narrow-band photometry, broad band optical colors, and  Gaia DR3 proper motions. We analyse the CaT region of 39 spectra with SNR $\geq$ 3, of stars distributed over a region reaching $\sim 8$ r$_{h}$ from the galaxy center. 
We combine our new sample with previously published  datasets, and can derive the most robust dynamical and metallicity properties of Boo~II to date.  In particular, we report the discovery of $9$ new member stars, including $6$ in common with B23. One additional star, apparently metal-rich, is  located at $\sim 4$ r$_{h}$ of the satellite. While it  has the right velocity, its $CaHK$ photometry places it onto the MW locus and it has no proper motion measurement. Future spectroscopic programs should confirm its membership.

All the confirmed members now have improved velocity measurements thanks to our FLAMES program and, more specifically, the use of the HR21 grating and with higher resolution than the spectroscopic setups previously used.
We derive the metallicity of 8 stars and double the number of stars with chemical information. We confirm the very metal-poor nature of the system.
Moreover, we provide the identification of two spectroscopically EMP stars in Boo~II. 

We find a systemic velocity of $\langle \mathrm{v} \rangle = -126.8^{+2.0}_{-1.5}$ km s$^{-1}$ and a dispersion of $5.6^{+1.8}_{-1.1}$ km s$^{-1}$. We note the possibility of a slight velocity gradient, but it is nevertheless still compatible with no velocity gradient at the $\sim 1.5\sigma$ level.

We also obtained spectroscopic data for two other UFDs, Leo~IV and Leo~V, but were only able to identify one new member which does not change our view of the two systems.

\newpage

\begin{acknowledgements}
This work has been carried out thanks to the support of the Swiss National Science Foundation.

Based on observations obtained with MegaPrime/MegaCam, a joint project of CFHT and CEA/DAPNIA, at the Canada-France-Hawaii Telescope (CFHT) which is operated by the National Research Council (NRC) of Canada, the Institut National des Science de l'Univers of the Centre National de la Recherche Scientifique (CNRS) of France, and the University of Hawaii.

The authors thank the International Space Science Institute, Bern, Switzerland for providing financial support and meeting facilities to the international team Pristine.

GB acknowledges support from the Agencia Estatal de Investigación del Ministerio de Ciencia en Innovación (AEI-MICIN) and the European Regional Development Fund (ERDF) under grant number PID2020-118778GB-I00/10.13039/50110001103 and the AEI under grant number CEX2019-000920-S

NFM gratefully acknowledge support from the French National Research Agency (ANR) funded project ``Pristine'' (ANR-18-CE31-0017) along with funding from the European Research Council (ERC) under the European Unions Horizon 2020 research and innovation programme (grant agreement No. 834148).

This work has made use of data from the European Space Agency (ESA) mission Gaia (https://www.cosmos. esa.int/gaia), processed by the Gaia Data Processing and Analysis Consortium (DPAC, https://www.cosmos.esa. int/web/gaia/dpac/consortium). Funding for the DPAC has been provided by national institutions, in particular the institutions participating in the Gaia Multilateral Agreement.

The Pan-STARRS1 Surveys (PS1) and the PS1 public science archive have been made possible through contributions by the Institute for Astronomy, the University of Hawaii, the Pan-STARRS Project Office, the Max-Planck Society and its participating institutes, the Max Planck Institute for Astronomy, Heidelberg and the Max Planck Institute for Extraterrestrial Physics, Garching, The Johns Hopkins University, Durham University, the University of Edinburgh, the Queen's University Belfast, the Harvard-Smithsonian Center for Astrophysics, the Las Cumbres Observatory Global Telescope Network Incorporated, the National Central University of Taiwan, the Space Telescope Science Institute, the National Aeronautics and Space Administration under Grant No. NNX08AR22G issued through the Planetary Science Division of the NASA Science Mission Directorate, the National Science Foundation Grant No. AST-1238877, the University of Maryland, Eotvos Lorand University (ELTE), the Los Alamos National Laboratory, and the Gordon and Betty Moore Foundation.

This project has received funding from the European Union's Horizon 2020 research and innovation programme under grant agreement No 730890. This material reflects only the authors views and the Commission is not liable for any use that may be made of the information contained therein.

Funding for the Sloan Digital Sky Survey V has been provided by the Alfred P. Sloan Foundation, the Heising-Simons Foundation, the National Science Foundation, and the Participating Institutions. SDSS acknowledges support and resources from the Center for High-Performance Computing at the University of Utah. The SDSS web site is \url{www.sdss.org}.

SDSS is managed by the Astrophysical Research Consortium for the Participating Institutions of the SDSS Collaboration, including the Carnegie Institution for Science, Chilean National Time Allocation Committee (CNTAC) ratified researchers, the Gotham Participation Group, Harvard University, Heidelberg University, The Johns Hopkins University, L’Ecole polytechnique fédérale de Lausanne (EPFL), Leibniz-Institut für Astrophysik Potsdam (AIP), Max-Planck-Institut für Astronomie (MPIA Heidelberg), Max-Planck-Institut für Extraterrestrische Physik (MPE), Nanjing University, National Astronomical Observatories of China (NAOC), New Mexico State University, The Ohio State University, Pennsylvania State University, Smithsonian Astrophysical Observatory, Space Telescope Science Institute (STScI), the Stellar Astrophysics Participation Group, Universidad Nacional Autónoma de México, University of Arizona, University of Colorado Boulder, University of Illinois at Urbana-Champaign, University of Toronto, University of Utah, University of Virginia, Yale University, and Yunnan University.

\end{acknowledgements}



\newpage

\begin{table*}

\caption{Properties of the new FLAMES spectroscopic sample. The individual spectroscopic metallicities for our calibration are reported only for members with S/N $\geq 10$. Confirmed new members are denoted with "Y" in the member column, while the "metal-rich candidate" is denoted with "MRC"}.

\setlength{\tabcolsep}{2.5pt}
\renewcommand{\arraystretch}{0.3}
\begin{sideways}
\begin{tabular}{cccccccccccccc}
\hline
RA (deg) & DEC (deg) & $g^\mathrm{SDSS}_0$ & $i^\mathrm{SDSS}_0$ & CaHK$_0$ & $\mathrm{v}_{r}$ (km/s) & $\mu_{\alpha}^{*}$ (mas.yr$^{-1}$) & $\mu_{\delta}$ (mas.yr$^{-1}$) &  S/N & [Fe/H]$_\mathrm{Spectro}$ & [Fe/H]$_\mathrm{Pristine}$ & Member \\

\hline

209.45242 & 12.80231 & 20.78 $\pm$ 0.03 & 20.40 $\pm$ 0.04 & 21.64 $\pm$ 0.13 & 152.8 $\pm$ 2.7 & $-$4.18 $\pm$ 1.28 & $-$1.19 $\pm$ 1.20 & 9.6 & $-$2.3 $\pm$ 0.1 & $-$99.0 &  N  \\ \\ 
209.44429 & 12.89236 & 20.26 $\pm$ 0.02 & 19.58 $\pm$ 0.02 & 21.18 $\pm$ 0.09 & $-$129.7 $\pm$ 0.6 & $-$2.48 $\pm$ 0.55 & 0.32 $\pm$ 0.44 & 18.3 & $-$2.5 $\pm$ 0.1 & $-$1.05 &  Y  \\ \\ 
209.63325 & 12.80231 & 19.49 $\pm$ 0.02 & 18.95 $\pm$ 0.02 & 20.05 $\pm$ 0.04 & 11.2 $\pm$ 1.3 & $-$5.70 $\pm$ 0.30 & $-$4.62 $\pm$ 0.23 & 26.0 & $-$2.8 $\pm$ 0.1 & $-$2.08 &  N  \\ \\ 
209.46717 & 12.74372 & 19.25 $\pm$ 0.01 & 18.97 $\pm$ 0.02 & 19.58 $\pm$ 0.03 & 230.4 $\pm$ 1.8 & 0.76 $\pm$ 0.32 & $-$0.21 $\pm$ 0.24 & 26.7 & $-$4.6 $\pm$ 0.5 & $-$2.02 &  N  \\ \\ 
209.5725 & 12.8035 & 20.74 $\pm$ 0.03 & 20.14 $\pm$ 0.03 & 21.52 $\pm$ 0.12 & $-$124.7 $\pm$ 0.5 & $-$2.60 $\pm$ 0.85 & 0.27 $\pm$ 0.81 & 14.0 & $-$2.1 $\pm$ 0.1 & $-$1.11 &  Y  \\ \\ 
209.65196 & 12.79147 & 20.13 $\pm$ 0.02 & 19.79 $\pm$ 0.03 & 20.59 $\pm$ 0.06 & $-$223.4 $\pm$ 1.4 & $-$1.85 $\pm$ 0.52 & $-$2.89 $\pm$ 0.42 & 18.2 & $-$3.0 $\pm$ 0.1 & $-$1.29 &  N  \\ \\ 
209.47608 & 13.03169 & 19.56 $\pm$ 0.02 & 19.29 $\pm$ 0.02 & 19.92 $\pm$ 0.03 & $-$149.6 $\pm$ 1.3 & 0.70 $\pm$ 0.51 & $-$0.59 $\pm$ 0.33 & 18.1 & $-$2.7 $\pm$ 0.1 & $-$1.41 &  N  \\ \\ 
209.354 & 12.80975 & 21.56 $\pm$ 0.05 & 20.89 $\pm$ 0.05 & 22.17 $\pm$ 0.21 & $-$62.7 $\pm$ 0.1 & --- & --- & 6.3 & --- & $-$2.41 &  N  \\ \\ 
209.50475 & 12.98253 & 21.29 $\pm$ 0.05 & 21.02 $\pm$ 0.07 & 22.14 $\pm$ 0.20 & 161.5 $\pm$ 3.0 & --- & --- & 6.1 & --- & $-$99.0 &  N  \\ \\ 
209.55775 & 12.98733 & 20.08 $\pm$ 0.02 & 19.73 $\pm$ 0.03 & 20.55 $\pm$ 0.05 & $-$71.5 $\pm$ 1.4 & $-$3.10 $\pm$ 0.50 & $-$2.54 $\pm$ 0.39 & 19.2 & $-$2.6 $\pm$ 0.1 & $-$1.35 &  N  \\ \\ 
209.51708 & 12.68917 & 20.42 $\pm$ 0.02 & 19.98 $\pm$ 0.03 & 20.90 $\pm$ 0.07 & 226.2 $\pm$ 1.7 & $-$8.06 $\pm$ 0.73 & $-$3.92 $\pm$ 0.57 & 13.1 & $-$3.1 $\pm$ 0.1 & $-$2.03 &  N  \\ \\ 
209.54108 & 12.79853 & 20.22 $\pm$ 0.02 & 19.57 $\pm$ 0.02 & 20.84 $\pm$ 0.07 & $-$128.5 $\pm$ 1.4 & $-$2.70 $\pm$ 0.55 & $-$0.71 $\pm$ 0.43 & 18.0 & $-$4.0 $\pm$ 0.4 & $-$2.35 &  Y  \\ \\ 
209.53246 & 12.82261 & 20.71 $\pm$ 0.03 & 20.09 $\pm$ 0.03 & 21.35 $\pm$ 0.10 & $-$126.0 $\pm$ 1.3 & $-$2.41 $\pm$ 0.70 & $-$0.05 $\pm$ 0.54 & 16.2 & $-$2.0 $\pm$ 0.1 & $-$2.06 &  Y  \\ \\ 
209.3935 & 12.88433 & 20.60 $\pm$ 0.03 & 20.12 $\pm$ 0.03 & 21.31 $\pm$ 0.10 & $-$2.2 $\pm$ 0.1 & $-$0.77 $\pm$ 0.98 & $-$3.03 $\pm$ 1.01 & 10.9 & --- & $-$0.80 &  N  \\ \\ 
209.560 & 12.91556 & 21.55 $\pm$ 0.06 & 20.89 $\pm$ 0.06 & 22.27 $\pm$ 0.21 & 79.6 $\pm$ 1.9 & --- & --- & 7.5 & $-$2.2 $\pm$ 0.1 & $-$1.87 &  N  \\ \\ 
209.70254 & 12.84258 & 20.51 $\pm$ 0.03 & 20.02 $\pm$ 0.03 & 21.10 $\pm$ 0.08 & 180.0 $\pm$ 0.7 & $-$1.61 $\pm$ 0.75 & $-$1.64 $\pm$ 0.63 & 11.1 & --- & $-$1.64 &  N  \\ \\ 
209.55479 & 12.83772 & 21.10 $\pm$ 0.04 & 20.42 $\pm$ 0.04 & 22.02 $\pm$ 0.17 & $-$111.3 $\pm$ 0.6 & $-$0.34 $\pm$ 1.15 & $-$1.92 $\pm$ 1.06 & 10.8 & $-$2.2 $\pm$ 0.1 & $-$0.93 &  Y  \\ \\ 
209.65579 & 12.72564 & 21.14 $\pm$ 0.04 & 20.80 $\pm$ 0.05 & 21.50 $\pm$ 0.12 & 220.8 $\pm$ 1.5 & --- & --- & 6.9 & --- & $-$2.12 &  N  \\ \\ 
209.46333 & 12.7540 & 21.31 $\pm$ 0.04 & 20.43 $\pm$ 0.04 & 22.20 $\pm$ 0.20 & $-$95.8 $\pm$ 1.0 & $-$5.73 $\pm$ 1.31 & 0.99 $\pm$ 1.91 & 11.3 & --- & $-$2.00 &  N  \\ \\ 
209.68962 & 12.87806 & 16.74 $\pm$ 0.01 & 15.88 $\pm$ 0.01 & 17.66 $\pm$ 0.01 & 156.4 $\pm$ 0.1 & $-$1.17 $\pm$ 0.06 & $-$0.90 $\pm$ 0.04 & 100.2 & --- & $-$1.82 &  N  \\ \\ 
209.47058 & 12.75111 & 20.28 $\pm$ 0.02 & 19.86 $\pm$ 0.03 & 20.95 $\pm$ 0.07 & 241.6 $\pm$ 2.0 & $-$4.14 $\pm$ 0.74 & $-$0.17 $\pm$ 0.46 & 14.5 & $-$1.8 $\pm$ 0.1 & $-$0.81 &  N  \\ \\ 
209.62492 & 12.94817 & 20.37 $\pm$ 0.03 & 19.70 $\pm$ 0.03 & 21.29 $\pm$ 0.09 & $-$118.2 $\pm$ 1.1 & $-$1.56 $\pm$ 0.56 & 0.28 $\pm$ 0.46 & 17.0 & $-$2.0 $\pm$ 0.1 & $-$0.88 &  Y  \\ \\ 
209.52746 & 12.69019 & 19.25 $\pm$ 0.01 & 18.94 $\pm$ 0.02 & 19.73 $\pm$ 0.03 & 137.9 $\pm$ 0.6 & 0.60 $\pm$ 0.27 & $-$7.25 $\pm$ 0.21 & 24.9 & $-$3.2 $\pm$ 0.1 & $-$0.99 &  N  \\ \\ 
209.40612 & 12.86347 & 18.15 $\pm$ 0.01 & 17.49 $\pm$ 0.01 & 18.83 $\pm$ 0.02 & $-$133.3 $\pm$ 0.2 & $-$2.48 $\pm$ 0.13 & $-$0.41 $\pm$ 0.09 & 48.8 & $-$3.5 $\pm$ 0.1 & $-$2.05 &  Y  \\ \\ 
209.505 & 12.99647 & 21.09 $\pm$ 0.04 & 20.65 $\pm$ 0.05 & 21.87 $\pm$ 0.16 & $-$42.9 $\pm$ 1.6 & 0.12 $\pm$ 1.75 & $-$4.05 $\pm$ 2.12 & 7.8 & $-$2.4 $\pm$ 0.1 & $-$0.47 &  N  \\ \\ 
209.43242 & 12.79056 & 20.21 $\pm$ 0.02 & 19.52 $\pm$ 0.02 & 20.80 $\pm$ 0.06 & $-$127.7 $\pm$ 0.8 & $-$2.38 $\pm$ 0.42 & $-$0.57 $\pm$ 0.35 & 19.2 & $-$2.9 $\pm$ 0.1 & $-$2.64 &  Y  \\ \\ 
209.65917 & 12.79481 & 20.26 $\pm$ 0.02 & 19.90 $\pm$ 0.03 & 20.83 $\pm$ 0.07 & 5.3 $\pm$ 1.2 & $-$3.11 $\pm$ 0.78 & $-$3.12 $\pm$ 0.59 & 18.9 & $-$2.3 $\pm$ 0.1 & $-$0.98 &  N  \\ \\ 
209.41867 & 12.92331 & 21.15 $\pm$ 0.04 & 20.53 $\pm$ 0.05 & 21.73 $\pm$ 0.14 & 96.3 $\pm$ 2.2 & 3.79 $\pm$ 1.64 & $-$5.74 $\pm$ 1.91 & 7.6 & $-$2.6 $\pm$ 0.1 & $-$2.35 &  N  \\ \\ 
209.49804 & 12.86472 & 20.61 $\pm$ 0.03 & 20.00 $\pm$ 0.03 & 21.30 $\pm$ 0.10 & $-$126.2 $\pm$ 1.3 & $-$1.11 $\pm$ 0.66 & $-$0.93 $\pm$ 0.51 & 16.3 & $-$2.6 $\pm$ 0.1 & $-$1.78 &  Y  \\ \\ 
209.465 & 13.02353 & 20.40 $\pm$ 0.03 & 19.98 $\pm$ 0.03 & 21.24 $\pm$ 0.10 & $-$106.0 $\pm$ 1.3 & $-$0.98 $\pm$ 0.74 & $-$2.21 $\pm$ 0.72 & 12.7 & $-$1.8 $\pm$ 0.1 & $-$99.0 &  N  \\ \\ 
209.37796 & 12.8160 & 21.50 $\pm$ 0.05 & 20.98 $\pm$ 0.06 & 22.08 $\pm$ 0.19 & 242.3 $\pm$ 0.6 & --- & --- & 6.2 & --- & $-$1.82 &  N  \\ \\ 
209.56796 & 12.93706 & 21.35 $\pm$ 0.05 & 20.70 $\pm$ 0.05 & 22.09 $\pm$ 0.18 & $-$197.5 $\pm$ 0.4 & --- & --- & 10.4 & $-$1.6 $\pm$ 0.1 & $-$1.66 &  N  \\ \\ 
209.48992 & 12.69061 & 21.41 $\pm$ 0.05 & 20.93 $\pm$ 0.06 & 22.24 $\pm$ 0.21 & $-$133.2 $\pm$ 1.1 & --- & --- & 6.0 & --- & $-$0.41 &  MRC  \\ \\ 
209.44946 & 12.79931 & 21.41 $\pm$ 0.05 & 20.95 $\pm$ 0.06 & 21.83 $\pm$ 0.15 & $-$59.5 $\pm$ 2.1 & --- & --- & 6.1 & --- & $-$2.55 &  N  \\ \\ 
209.36496 & 12.77058 & 21.62 $\pm$ 0.06 & 21.33 $\pm$ 0.08 & 23.06 $\pm$ 0.41 & 68.8 $\pm$ 0.3 & --- & --- & 4.7 & --- & $-$99.0 &  N  \\ \\ 
209.42846 & 12.85108 & 16.77 $\pm$ 0.01 & 16.55 $\pm$ 0.01 & 17.19 $\pm$ 0.01 & $-$109.1 $\pm$ 0.3 & $-$6.84 $\pm$ 0.07 & $-$14.6 $\pm$ 0.05 & 81.7 & $-$4.4 $\pm$ 0.2 & $-$0.51 &  N  \\ \\ 
209.47825 & 12.75947 & 21.10 $\pm$ 0.04 & 20.66 $\pm$ 0.05 & 21.51 $\pm$ 0.11 & 146.5 $\pm$ 1.9 & $-$1.14 $\pm$ 1.75 & 1.63 $\pm$ 2.18 & 6.9 & --- & $-$2.50 &  N  \\ \\ 
209.44692 & 12.84725 & 21.03 $\pm$ 0.04 & 20.28 $\pm$ 0.03 & 21.93 $\pm$ 0.17 & $-$31.2 $\pm$ 2.3 & $-$2.09 $\pm$ 1.07 & 1.17 $\pm$ 1.13 & 12.2 & $-$2.0 $\pm$ 0.1 & $-$1.38 &  N  \\ \\ 
209.583 & 12.88564 & 17.91 $\pm$ 0.01 & 17.65 $\pm$ 0.01 & 18.38 $\pm$ 0.01 & 15.2 $\pm$ 0.3 & $-$1.68 $\pm$ 0.12 & $-$12.67 $\pm$ 0.10 & 43.5 & $-$3.3 $\pm$ 0.1 & $-$0.67 &  N  \\ \\

\end{tabular}
\end{sideways}
\end{table*}

\newpage

\begin{table*}

\caption{Probable misidentified members from the literature (K09 and B23). The mention ``CaHK'' in the last column indicates that the $CaHK$ magnitude was decisive in the decision-making.}

\setlength{\tabcolsep}{2.5pt}
\renewcommand{\arraystretch}{0.3}
\begin{tabular}{cccccccccccccc}
\hline
RA (deg) & DEC (deg) & $g^\mathrm{SDSS}_0$ & $i^\mathrm{SDSS}_0$ & $CaHK_0$ & $\mathrm{v}_{r} (km/s)$ & $\mu_{\alpha}^{*}$ (mas.yr$^{-1}$) & $\mu_{\delta}$ (mas.yr$^{-1}$) & [Fe/H]$_\mathrm{spectro}$  \\

\hline

209.53947 & 12.85719 & 20.37 $\pm$ 0.03 & 19.74 $\pm$ 0.03 & 20.94 $\pm$ 0.07 & $-$138.5 $\pm$ 2.6 & $-$2.71 $\pm$ 0.53 & 0.02 $\pm$ 0.44 & --- & CaHK \\ \\ 
209.5729 & 12.86183 & 20.39 $\pm$ 0.03 & 19.78 $\pm$ 0.03 & 21.04 $\pm$ 0.08 & $-$135.8 $\pm$ 2.4 & $-$3.24 $\pm$ 0.53 & $-$0.47 $\pm$ 0.42 & --- & CaHK \\ \\ 
209.52933 & 12.85634 & 18.58 $\pm$ 0.01 & 18.34 $\pm$ 0.01 & 18.30 $\pm$ 0.01 & $-$118.4 $\pm$ 1.4 & $-$2.59 $\pm$ 0.16 & $-$0.61 $\pm$ 0.12 & --- & CaHK \\ \\ 

\end{tabular}
\end{table*}

\newpage

\begin{table*}

\caption{CMD (p$_{\mathrm{CMD}}$), radial velocity (p$_{\mathrm{v}}$) and proper motion (p$_{\mathrm{PM}}$) membership probabilities of stars in the FLAMES sample. The member with a p$_{\mathrm{CMD}}$ denoted "HB" is the star consistent with being a HB star. Since our CMD membership computation method does not take into account a HB model, no CMD membership is available for such stars.}

\setlength{\tabcolsep}{2.5pt}
\renewcommand{\arraystretch}{0.3}
\begin{tabular}{cccccccccccccc}
\hline
RA (deg) & DEC (deg) & p$_{\mathrm{CMD}}$ & p$_{\mathrm{v}}$ & p$_{\mathrm{PM}}$ & Member \\

\hline

209.45242 & 12.80231 & 0.00037 & 0.00000 & 0.83464 &  N  \\ \\ 
209.44429 & 12.89236 & 0.65840 & 0.98227 & 0.97550 &  Y  \\ \\ 
209.63325 & 12.80231 & 0.00000 & 0.00000 & 0.00000 &  N  \\ \\ 
209.46717 & 12.74372 & 0.00000 & 0.00000 & 0.00000 &  N  \\ \\ 
209.5725 & 12.8035 & 0.53759 & 0.98161 & 0.96161 &  Y  \\ \\ 
209.65196 & 12.79147 & 0.00000 & 0.00000 & 0.47323 &  N  \\ \\ 
209.47608 & 13.03169 & 0.00000 & 0.72769 & 0.00000 &  N  \\ \\ 
209.354 & 12.80975 & 0.10807 & 1e-0500 & --- &  N  \\ \\ 
209.50475 & 12.98253 & 0.50323 & 0.00000 & --- &  N  \\ \\ 
209.55775 & 12.98733 & 0.00000 & 0.00745 & 0.44630 &  N  \\ \\ 
209.51708 & 12.68917 & 0.00000 & 0.00000 & 0.00000 &  N  \\ \\ 
209.54108 & 12.79853 & 0.66842 & 0.98154 & 0.97632 &  Y  \\ \\ 
209.53246 & 12.82261 & 0.63008 & 0.98278 & 0.97322 &  Y  \\ \\ 
209.3935 & 12.88433 & 0.00195 & 0.00000 & 0.50802 &  N  \\ \\ 
209.560 & 12.91556 & 0.16747 & 0.00000 & --- &  N  \\ \\ 
209.70254 & 12.84258 & 0.00144 & 0.00000 & 0.88547 &  N  \\ \\ 
209.55479 & 12.83772 & 0.41794 & 0.84146 & 0.64283 &  Y  \\ \\ 
209.65579 & 12.72564 & 0.17370 & 0.00000 & --- &  N  \\ \\ 
209.46333 & 12.7540 & 2e-0500 & 0.04141 & 0.55894 &  N  \\ \\ 
209.68962 & 12.87806 & 0.00000 & 0.00000 & 1e-0500 &  N  \\ \\ 
209.47058 & 12.75111 & 0.00000 & 0.00000 & 0.77016 &  N  \\ \\ 
209.62492 & 12.94817 & 0.64702 & 0.97377 & 0.86422 &  Y  \\ \\ 
209.52746 & 12.69019 & 0.00000 & 0.00000 & 0.00000 &  N  \\ \\ 
209.40612 & 12.86347 & 0.00000 & 0.98042 & 0.99485 &  Y  \\ \\ 
209.505 & 12.99647 & 0.18890 & 0.00012 & 0.43439 &  N  \\ \\ 
209.43242 & 12.79056 & 0.64358 & 0.98066 & 0.98482 &  Y  \\ \\ 
209.65917 & 12.79481 & 0.00000 & 0.00000 & 0.39250 &  N  \\ \\ 
209.41867 & 12.92331 & 0.52219 & 0.00000 & 0.00944 &  N  \\ \\ 
209.49804 & 12.86472 & 0.53114 & 0.98275 & 0.78679 &  Y  \\ \\ 
209.465 & 13.02353 & 0.00000 & 0.71273 & 0.60172 &  N  \\ \\ 
209.37796 & 12.8160 & 0.39795 & 0.00000 & --- &  N  \\ \\ 
209.56796 & 12.93706 & 0.40436 & 1e-0500 & --- &  N  \\ \\ 
209.48992 & 12.69061 & 0.51039 & 0.97533 & --- &  N  \\ \\ 
209.44946 & 12.79931 & 0.52035 & 0.00000 & --- &  N  \\ \\ 
209.36496 & 12.77058 & 0.71876 & 0.00000 & --- &  N  \\ \\ 
209.42846 & 12.85108 & 0.00000 & 0.54348 & 0.00000 &  N  \\ \\ 
209.47825 & 12.75947 & 0.20659 & 0.00000 & 0.79975 &  N  \\ \\ 
209.44692 & 12.84725 & 0.10727 & 0.00000 & 0.91071 &  N  \\ \\ 
209.583 & 12.88564 & 0.00000 & 0.00000 & 0.00000 &  N  \\ \\

\end{tabular}
\end{table*}

\end{document}